\documentclass[a4paper]{spie}  
\usepackage[right=1.925 cm, left=1.925 cm, top=2.54 cm, bottom=4.94 cm]{geometry}

\usepackage{amsmath,amsfonts,amssymb}
\usepackage{graphicx}
\usepackage{amsmath,amssymb,bm}
\usepackage{graphicx}
\usepackage{float}
\usepackage{color,dsfont}
\usepackage{braket}
\usepackage{mathtools}

\DeclareMathOperator{\sech}{sech}

\title{Long-distance quantum key distribution using concatenated entanglement swapping with practical resources}

\author[a,b,c]{Aeysha Khalique}
\author[b,c,d,e]{Barry C. Sanders}
\affil
[a]{School of Natural Sciences, National University of Sciences and Technology,
	H-12 Islamabad, Pakistan}
\affil
[b]{Shanghai Branch, CAS Center for Excellence and Synergetic Innovation Center in Quantum Information and Quantum Physics, University of Science and Technology of China, Shanghai 201315, China}
\affil
[c]{Hefei National Laboratory for Physical Sciences at Microscale, University of Science and Technology of China, Hefei, Anhui 230026, China}
\affil
[d]{Institute for Quantum Science and Technology, 
	University of Calgary, Alberta T2N 1N4, Canada}
\affil
[e]{Program in Quantum Information Science,
	Canadian Institute for Advanced Research, Toronto, Ontario M5G 1Z8, Canada}

\authorinfo{Further author information: (Send correspondence to A.K.)\\A.K.: E-mail: ashkhalique@gmail.com
\\  B.C.S.: E-mail: sandersb@ucalgary.ca
}
\begin{document} 
\maketitle

\begin{abstract}
We present our approach for sharing photons and assessing resultant four-photon visibility between two distant parties using concatenated entanglement swapping. In addition we determine the corresponding key generation rate and the quantum bit-error rate. Our model is based on practical limitations of resources, including multipair parametric down-conversion sources, inefficient detectors with dark counts and lossy channels. Through this approach, we have found that a trade-off is needed between experimental run-time, pair-production rate and detector efficiency. Concatenated entanglement swapping enables huge distances for quantum key-distribution but at the expense of low key generation rate.
\end{abstract}


\section{INTRODUCTION}
\label{sec:intro}  
Quantum communication provides a means for secure communication in open channels~\cite{GRT+02}. One of the primary goals of quantum communication is to develop the ability to communicate to arbitrary distances. Experimentally communication distances have been limited to a few hundred kilometres. Recently quantum key distribution (QKD) of up to 200~km has been achieved with Measurement-Device-Independent QKD (MDI-QKD)~\cite{TYC+14}. A distance of upto~250~km has been achieved by using subcarrier wave modulation method which employs the Bennet-Brassard (BB84) protocol~\cite{GEN+16}. Quantum relays and quantum repeaters are promising setups to achieve the ultimate goal of long-distance quantum communication~\cite{GT07}. In principle, any distance is achievable by using quantum relays. In practice, however, the allowed distance is limited by resource imperfections. These imperfections also limit the key-generation rate which will affect the efficacy of the system.

Quantum relays and repeaters have been investigated for long-distance key distribution~\cite{KGD+16}, which models the resources with approximations. We provide a rigorous model for a quantum relay setup based on entanglement swapping, in which we have included the imperfections of the sources, the channel and the detectors~\cite{KTS13, KS14}. This concatenated entanglement swapping setup is then extended to key distribution protocol~\cite{KS15}, which relies on Bennett-Brassard-Mermin (BBM92)~\cite{BBM92} protocol. In this paper we review our approach to model concatenated entanglement swapping and key distribution protocol based on that. This approach acts as a base model for quantum memories, and by extension for quantum repeaters. 

This paper is organized as follows. In Sec.~\ref{sec:swappingresources}, we explain the entanglement swapping process and discuss the resources proposed for an experimental setup. In Sec.~\ref{sec:singleswap}, we present the model for a single swap and calculation of four photon visibility based on that. The concatenated entanglement swapping setup for long-distance quantum communication and the corresponding results for visibility are shown in Sec.~\ref{sec:nswaps}. In Sec.~\ref{sec:qkd}, we present the QKD protocol based on concatenated entanglement swapping and show the results for maximum key generation rates with optimized resource parameters. Finally, we conclude in Sec.~\ref{sec:conclusions}.

\section{Resources in entanglement swapping}
\label{sec:swappingresources}
The achievable distance in quantum communication can be increased by entanglement swapping. In this section we briefly review the entanglement swapping procedure and the practical resources used in a swapping experiment. In Sec.~\ref{sec:qswapping}, we explain the entanglement swapping procedure and we explain the resources in Sec.~\ref{sec:resources}.

\subsection{Entanglement Swapping}\label{sec:qswapping}
\begin{figure}[ht]
\begin{center}
\includegraphics[scale=0.5]{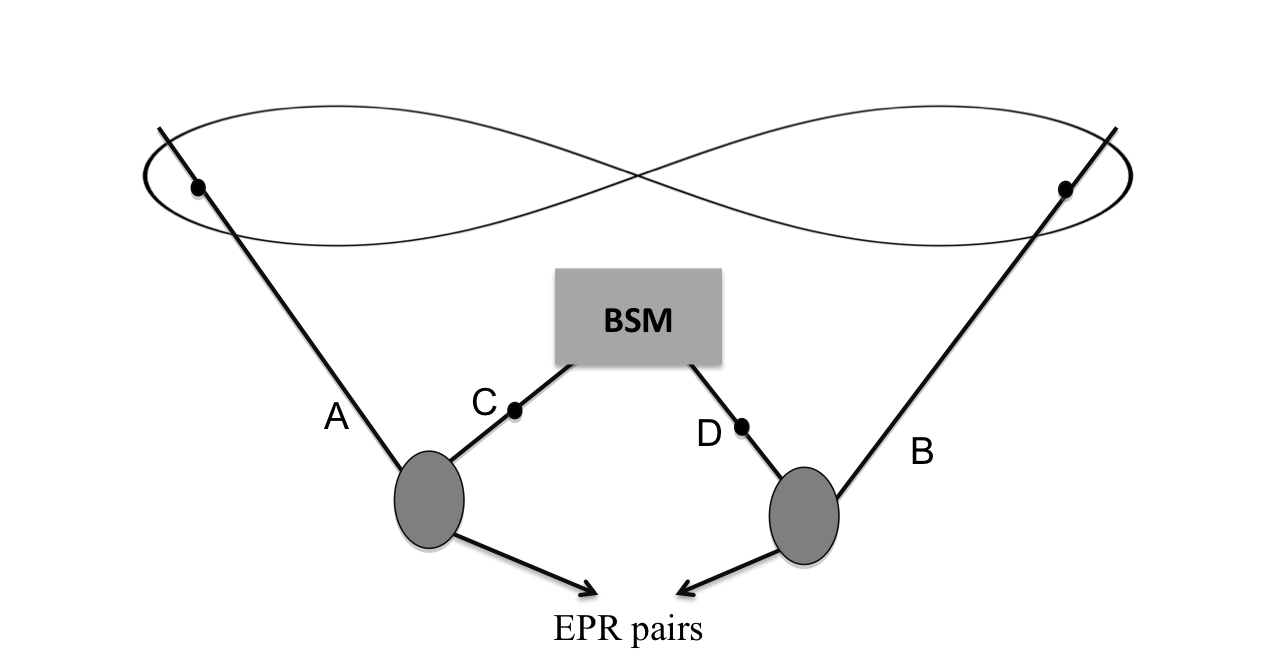}
\caption{Two Einstein-Podolsky-Rosen (EPR) sources produce entangled pairs AC and BD. A Bell State Measurement (BSM) on C and D entangles A and B.
}
\label{fig:qswapping}
\end{center}
\end{figure}

 Entanglement swapping provides a means for entangling distant parties who have never interacted in the past. Fig.~\ref{fig:qswapping} shows the entanglement of possibly distant parties A and B, when each of their entangled partners, C and D, undergo a Bell State Measurement (BSM). BSM distinguishes between the four Bell states
 \begin{align}
 \ket{\psi^+}&=\frac{1}{\sqrt 2}\left(\ket{\text {HV}}+\ket{\text{VH}}\right);\hspace{0.2cm} \ket{\psi^-}=\frac{1}{\sqrt 2}\left( \ket{\text{HV}}-\ket{\text{VH}}\right);\nonumber\\
 \ket{\phi^+}&=\frac{1}{\sqrt 2}\left(\ket{\text{HH}}+\ket{\text{VV}}\right);\hspace{0.2cm} \ket{\phi^-}=\frac{1}{\sqrt 2}\left(\ket{\text{HH}}-\ket{\text{VV}}\right).
 \end{align}
Entanglement swapping is evident from the fact that the combined entangled state of AC and BD can be written as
\begin{align}
\Ket {\psi ^+}_{\rm {AC}} \Ket {\psi ^+}_{\rm {BD}}=&\frac{1}{2}\left[\Ket {\psi ^+}_{\rm {AB}}\Ket {\psi ^+}_{\rm {CD}}+\Ket {\psi ^-}_{\rm {AB}}\Ket {\psi ^-}_{\rm {CD}}+\Ket {\phi ^+} _{\rm {AB}} \Ket {\phi ^+}_{\rm {CD}}+\Ket {\phi ^-}_{\rm {AB}} \Ket {\phi ^-}_{\rm {CD}}\right].
\end{align}
Thus BSM on C and D projects A and B into the corresponding Bell state.
\subsection{Resources}\label{sec:resources}
In a typical entanglement swapping setup the resources of concern are the entanglement source, the channels and the detectors. All these resources have imperfections. We consider a parametric down-conversion entanglement source, which produces multipairs of entangled photons. The state of the photons entangled in horizontal-vertical (H-V) polarization is
\begin{align}
	\ket{\chi}
		&=\text{e}^{i\chi\left(\hat a_{\rm H}^\dagger \hat c_{\rm H}^\dagger+\hat a_{\rm V}^\dagger
			\hat c_{\rm V}^\dagger+{\rm hc}\right)}\ket{\rm{vac}}\nonumber\\
		&=\sech^2\chi\text{e}^{i\tanh\chi\left(\hat a_{\rm H}^\dagger \hat c_{\rm H}^\dagger+\hat a_{\rm V}^\dagger \hat c_{\rm V}^\dagger\right)}\ket{\rm{vac}},
\end{align}
where, $\chi$ is the multipair production rate of the source and $\hat a_{\rm H}^\dagger$ and $\hat c_{\rm H}^\dagger$ are the creation operator for horizontally polarized photon in spatial mode $A$ and $C$ respectively. $\hat a_{\rm V}^\dagger$ and $\hat c_{\rm V}^\dagger$ are the corresponding ones for vertically polarized photons.

We consider a fibre optic channel with distance-dependent loss coefficient $\alpha$. The channel efficiency is given as
\begin{equation}
\eta_{\rm t}=\text{e}^{-\left(\alpha \ell +\alpha_0\right)/10},\nonumber
\end{equation}
where $l$ is the length of the fibre and $\alpha_0$ is the distance independent loss. The same model can be employed to model free space transmission.

\begin{figure}[ht]
\begin{center}
\includegraphics[scale=0.6]{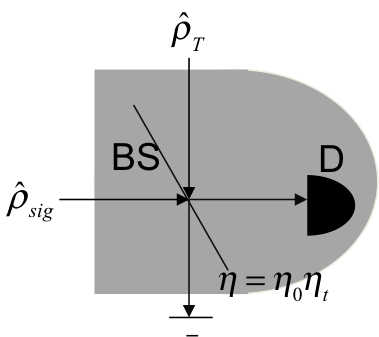}
\caption{The imperfect detector is replaced by a beam splitter (BS) and a perfect detector (D). On one port of BS, signal beam $\rho_{\rm{sig}}$ is incident and on the other port stray photons $\rho_{\rm{T}}$ are incident. The beam splitter has efficiency $\eta$, which comprises detector's intrinsic efficiency $\eta_0$ and channel efficiency $\eta_t$.   
}
\end{center}
\label{fig:detector}
\end{figure}

The `real' detectors are modelled by pairing perfect detectors with a beam splitters~\cite{RR06}, as shown in Fig.~\ref{fig:detector}.
Both the detector's intrinsic efficiency, $\eta_0$ and the channel transmission efficiency, $\eta_t$ are included in the transmission efficiency of the beam splitter, which in turn is $\eta=\eta_0 \eta_t$. The dark counts of the detector are modelled by a thermal source of light which represents stray photons and are incident on one port of the beam splitter. These photons at dimensionless temperature $T=\text{e}^{2r}-1$ are described by the density matrix~\cite{RR06}
\begin{equation}
\hat\rho_T=\frac{4(T+1)T}{T^2}\sum_{n=0}^{\infty}\left(\frac{T}{T+2}\right)^{2n}\ket n\bra n.
\end{equation}
Here $n$ is the photon number state and appropriate choice of $r$ models the mean photon number and dark count probability. The signal photons with density matrix~$\rho_{\rm{sig}}$ is incident on the other port. Threshold detectors have two possibilities, with $q=0$ representing no photon detected and $q=1$ representing a click. The probability of detecting $q$ photons when $i$ photons are incident is
\begin{align}
P\left(q=0|i\right)&=\left(1-\wp\right)\left[1-\eta\left(1-\wp\right)\right]^i,\\
&=1-P\left(q=1|i\right),
\end{align}
where, $i$ represents the number of photons in the signal $\hat\rho_{\rm{sig}}=\ket i\bra i$ and $\wp$ is the dark count probability. The detectors are independent of each other and the conditional probability of detecting $q$, $r$, $s$ and $t$ photons, each on one of four detectors when $i$, $j$, $k$ and $l$ signal photons are incident, respectively, is the following product of four probabilities
 \begin{align}
P\left(qrst|ijkl\right)&=P(q|i)P(r|j)P(s|k)P(t|l).
\label{eq:probproduct}
\end{align} 
This detector model is applicable to all quadruples of detectors.

\section{Practical single swap: coincidence probabilities and visibility}
\label{sec:singleswap}
\begin{figure}[ht]
\begin{center}
\includegraphics[scale=0.6]{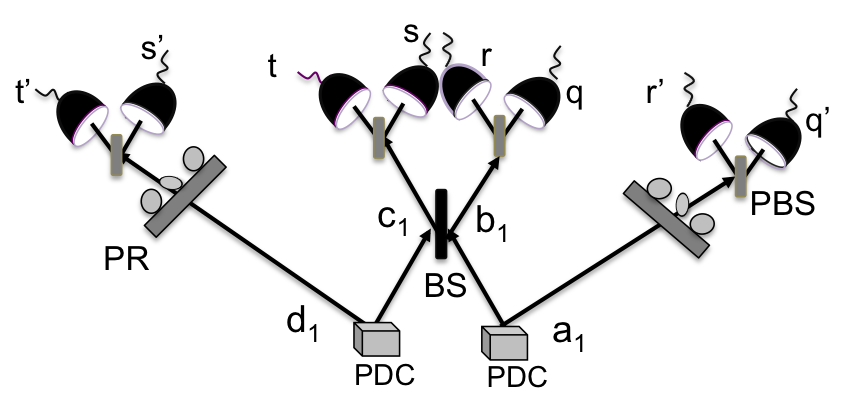}
\caption{The experimental setup for single swap.  $q'$ and $r'$ represent the clicks on detectors at A and $s'$ and $t'$ are clicks on those at B. $q$, $r$, $s$ and $t$ are the clicks at inner detector. PR are the polarizer rotators, PBS are polarization beam splitters and $BS$ is the beam splitters. PDC are parametric down-conversion sources.
}
\label{fig:singleswap}
\end{center}
\end{figure}
In practice, entanglement swapping setup consists of two parametric down-conversion (PDC) sources. Figure~\ref{fig:singleswap} shows two parties A and B, which get entangled by BSM at the two inner ports. The BSM setup consists of a beam splitter followed by polarization rotators and polarization beam splitters, which separate the horizontal and vertical polarized photons. These photons are then detected on the four photo-detectors. The detector clicks corresponding to ideal BSM for various Bell states are given in~Table~\ref{table:bsm}.
\begin{table}
\caption{The entangled state of the two parties corresponding to different clicks in the four detectors.}
\begin{center}
\begin{tabular}{|l|l|l|}
\hline
 Bell state&$(qrst)$ \\ \hline
$\ket{\psi^+}$&$(1010)\lor(0101)$ \\ \hline
$\ket{\psi^-}$&$(0110)\lor(1001)$ \\ \hline
$\ket{\phi^\pm}$& $(2000)\lor(0200)\lor(0020)\lor(0002)$\\ \hline
\end{tabular}
\label{table:bsm}
\end{center}
\end{table}

The practical resources do not lead to a perfect Bell state at the outer detectors with A and B. The maximum conditional probability of detecting clicks, $Q_{\rm{max}}(qrst)=\underset{q'r's't'}{\rm{max}}Q(q'r's't'|qrst;\chi,\wp,\eta)$, given the detector clicks at the four inner detectors mentioned in Table~\ref{table:bsm}, then depends on the resource parameters, $\chi$, $\wp$ and $\eta$. The same is true for minimum coincidence. The degree of entanglement is then quantified by using visibility, $V$, defined as
\begin{align} 
V\left(\chi,\wp,\eta\right)&=\frac{Q_{\rm{max}}-Q_{\rm{min}}}{Q_{\rm{max}}+Q_{\rm{min}}}
\label{eq:visibility}
\end{align}

Conditional probability $Q$ is calculated by the following course of action on the photons produced by the two PDCs~\cite{SHST09}. The photons in the inner two channels undergo the action of beam splitter $U_{\rm B}$ yielding $\ket{\Xi}=U_{\rm{B}}{\ket \chi}_\text{AB}{\ket \chi}_\text{CD}$ and the ideal detection of photons $i$, $j$, $k$ and $l$ at the inner detector is reflected by Fock projection $\Pi^{\rm{inn}}_{ijkl}$ which yields the state 
\begin{align}
\ket{\tilde {\Xi}}^\text{out}_{ijkl}:=\frac{\bra{ijkl}_{\rm{BC}}\Pi^{\rm{inn}}_{ijkl}{\ket \Xi}}{\sqrt{P(ijkl)}} 
\end{align}
at the outer ports. Here, $P(ijkl)=\braket{\Xi|\Pi^{\rm{inn}}_{ijkl}|\Xi}$. The noisy detection at the inner ports produces a mixed state $\rho_{qrst}^{\text{out}}$ at the outer ports, 
\begin{align}
\rho_{qrst}^{\text{out}} = \sum P(ijkl|qrst) \ket{\tilde\Xi}^\text{out}_{ijkl}\!\bra{\tilde\Xi},
\end{align}
where, $P(ijkl|qrst)$ is the conditional probability that $ijkl$ photons are detected ideally given actual detection $qrst$. This probability can be found from the known probability $P(qrst|ijkl)$ given in Eq.~(\ref{eq:probproduct}) by using Bayes Theorem
\begin{align}
 P(ijkl|qrst)&=\frac{P(qrst|ijkl)P(ijkl)}{P(qrst)}\\
&=P(q|i)P(r|j)P(s|k)P(t|l)P(ijkl)/P(qrst).
\end{align}
The conditional probability of detecting, with ideal detectors, $i'j'k'l'$ photons at the outer detectors given actual counts $qrst$ at the inner ones after passing through the polarization rotators at angles $\delta_A$ and $\delta_B$ is given by,
\begin{align}
& P(i'j'k'l'|qrst)=\bra{i'j'k'l'} U(\delta_A)U(\delta_B)\rho_{qrst}^{\text{out}} U^\dagger(\delta_A)U^\dagger(\delta_B)\ket{i'j'k'l'}.
\end{align}
The coincidence probability $Q$ of detecting actual photons $q'r's't'$ at the outer four detectors given $qrst$ at the inner ones is given by
\begin{align}
Q(q'r's't'|qrst)=\sum_{i'j'k'l'} P(q'r's't'|i'j'k'l';\wp,\eta)P(i'j'k'l'|qrst;\chi,\wp,\eta),
\end{align}
The $Q$ contains empirical parameters and can be calculated.

\section{Extending the distance by arbitrary swaps}
\label{sec:nswaps}
Communication distance can be extended by considering the concatenation of signal swaps. We analyze such a setup and calculate the corresponding visibility. We give a closed form solution for calculation of $Q_{\rm{ext}}(qrst)$ for an arbitrary number of swaps in Sec.~\ref{sec:cfsnswaps}. In Sec.~\ref{sec:n3visibility}, we give our results for calculation of visibility for $N\le3$.

\subsection{Closed form solution for calculation of $Q_{\rm{ext}}(qrst)$ for $N$ swaps}
\label{sec:cfsnswaps}
\begin{figure}[ht]
\begin{center}
\includegraphics[scale=0.6]{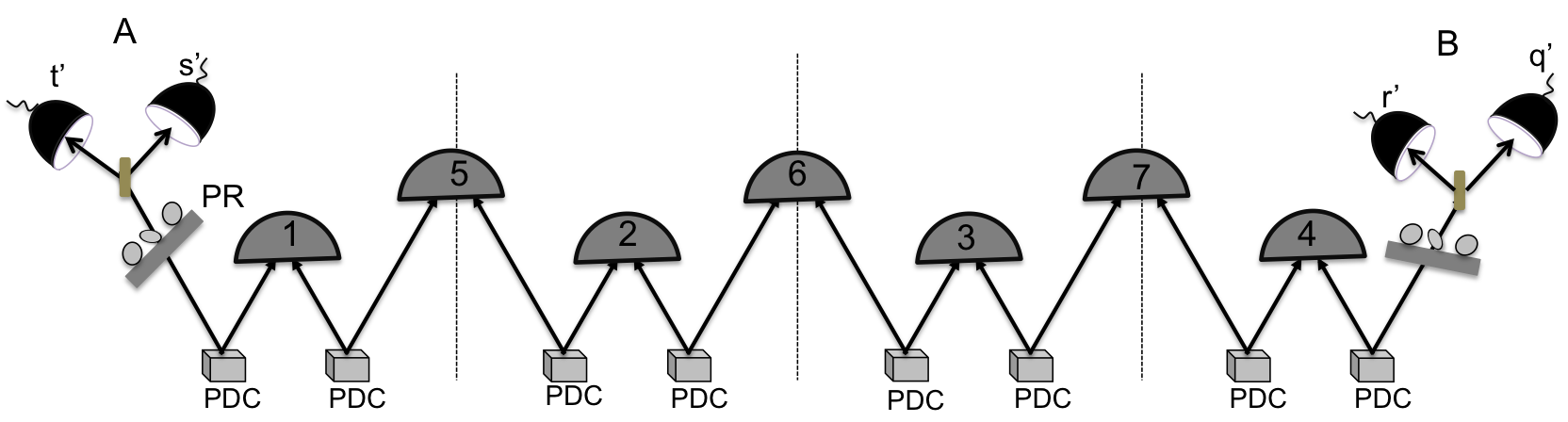}
\caption{Setup for concatenated entanglement swapping. For four concatenated swaps, seven BSMs  at the inner stations, entangle A and B at the two outer ends.
}
\label{fig:nswaps}
\end{center}
\end{figure}

The configuration for concatenated $N$ swaps is shown in Fig.~\ref{fig:nswaps}.
For $N$ swaps there are $2N-1$ BSMs. Ideally, successful BSM at the inner stations entangles distant parties A and B at the extreme ends. However, practically, the maximum probability of clicks at the outer detectors corresponding to clicks $\bm{qrst}$ at the inner ones is dependent on resource parameters $\chi$, $\wp$ and $\eta$,
\begin{equation}
Q_{\rm{ext}}(qrst)=\underset{q'r's't'}{\rm{ext}}Q(q'r's't'|\bm{qrst};\chi,\wp,\eta), 
\end{equation}  
where, $\bm q=\{q_1,q_2,\dots,q_{2N-1}\}$ and the same goes for $\bm r$, $\bm s$ and $\bm t$. 

The closed form solution of the conditional probability $P\left(i'j'k'l'|\bm{qrst}\right)$ is given as~\cite{KS14} 
\begin{align}
&P\left(i'j'k'l'|\bm{qrst}\right)= \sum_{\bm{ijkl}}P\left(\bm{ijkl}|\bm{qrst}\right)
\bra{i'j'k'l'} U(\alpha)U\left(\delta\right)\ket{\tilde\Xi}^\text{out}_{\bm{ijkl}}\!\bra {\tilde\Xi}U^\dagger(\alpha)U^\dagger(\delta)\ket{i'j'k'l'}\nonumber\\
&=
  \sum_{\bm{ijkl}}\frac{P\left(\bm{qrst}|\bm{ijkl}\right)}{P\left(\bm{qrst}\right)}
  \Bigg(
  \frac{1}{\sqrt{2^{i_1+j_1+k_1+l_1}i_1!j_1!k_1!l_1!}}\frac{\left(\tanh\chi\right)^{i_1+j_1+k_1+l_1}}       
         {\cosh^{4N}\chi}
        \sum_{\mu_1=0}^{i_1}\sum_{\nu_1=0}^{j_1}\sum_{\kappa_1=0}^{k_1}
         \sum_{\lambda_1=0}^{l_1}
            (-1)^{\mu_1+\nu_1}
           {i_1\choose \mu_1}{j_1\choose \nu_1}\nonumber\\
           &\times{k_1\choose \kappa_1}{l_1
                \choose \lambda_1}\dots
                \frac{1}{\sqrt{2^{i_{2N-1}+j_{2N-1}+k_{2N-1}+l_{2N-1}}i_{2N-1}!j_{2N-1}!k_{2N-1}!l_{2N-1}!}}\frac{\left(\tanh\chi\right)^{i_{2N-1}+j_{2N-1}+k_{2N-1}+l_{2N-1}}} {\cosh^{4N}\chi}\nonumber\\
        &\times \sum_{\mu_{2N-1}=0}^{i_{2N-1}}\sum_{\nu_{2N-1}=0}^{j_{2N-1}}\sum_{\kappa_{2N-1}=0}^{k_{2N-1}}
         \sum_{\lambda_{2N-1}=0}^{l_{2N-1}}
            (-1)^{\mu_{2N-1}+\nu_{2N-1}}
           {i_p\choose \mu_{2N-1}}{j_{2N-1}\choose \nu_{2N-1}}{k_{2N-1}\choose \kappa_{2N-1}}{l_{2N-1}
                \choose \lambda_{2N-1}}\Bigg)\nonumber\\
     &\times\prod_{n=1}^{N-1}
        \Omega\left(\mu_{n},\lambda_{n},i_{N+n},l_{N+n}\right)
            \Omega\left(\nu_{n},\kappa_{n},j_{N+n},k_{N+n}\right)
       \frac{\sqrt{i_{N+n}!j_{N+n}!k_{N+n}!l_{N+n}!}}       
        {\sqrt{2}^{i_{N+n}+j_{N+n}+k_{N+n}+l_{N+n}}}\nonumber\\
    &\times\delta_{i_{N+n}+l_{N+n},    
        \mu_{n}+\lambda_{n}+i_{n+1}+l_{n+1}-
         \mu_{n+1}-\lambda_{n+1}}
         \delta_{j_{N+n}+k_{N+n},
          \nu_{n}+\kappa_{n}+j_{n+1}+k_{n+1}-
            \nu_{n+1}-\kappa_{n+1}}\nonumber\\
    &\times(\nu_N+\kappa_N)!\left(j_1+k_1-\nu_1-\kappa_1\right)!\sqrt{\frac{j'!k'!}{i'!l'!}}
         \sum_{n_a=0}^{\text{min}\left [j',\nu_N+\kappa_N\right ]}
            \sum_{n_d=0}^{\text{min}\left [k',j_1+k_1-\nu_1-\kappa_1\right]}\nonumber\\
    &\times
             \left(\rm{i}\tan\frac{\delta_{\text {A}}}{2}\right)^{\nu_N+\kappa_N+j'-2n_a}
    \left(\cos\frac{\delta_{\text A}}
           {2}\right)^{i'+j'-2n_a}\left(\rm i\tan\frac{\delta_{\text B}}{2}\right)^{k'+j_1+k_1-\nu_1-\kappa_1-2n_d}
              \left(\cos\frac{\delta_{\text B}}{2}\right)^{l'+k'-2n_d}
          \nonumber\\
   &\times\frac{(i'+j'-n_a)!(l'+k'-nd)!}{n_a!n_d!\left(j'-n_a\right)!\left(k'-n_d\right)!\left(\nu_N+\kappa_N-n_a\right)! 
           \left(j_1+k_1-\nu_1-\kappa_1-n_d\right)!}\nonumber\\
   &\times\delta_{i'+j',
         \mu_N+\nu_N+\kappa_N+\lambda_N}
\delta_{k'+l',i_1+j_1+k_1+l_1-\mu_1-\nu_1-\kappa_1-\lambda_1}.
\end{align}
Here,
\begin{align}
\label{eq:Omega}
	\Omega(\mu_n,
            \lambda_n,i_{N+n},l_{N+n})
		=&\sum_{\gamma=0}^{\mu_n+\lambda_n}{\mu_n+\lambda_n\choose \gamma}
		{{i_{N+n}+l_{N+n}-\mu_n-\lambda_n }\choose {i_{N+n}-\gamma}}
		(-1)^{\mu_n+\lambda_n-\gamma}
\end{align}
is the factor resulting from BSM connecting the adjacent swaps.

\subsection{Visibility for $N\le3$ swaps}
\label{sec:n3visibility}
\begin{figure}[ht]
\begin{center}
\includegraphics[scale=0.6]{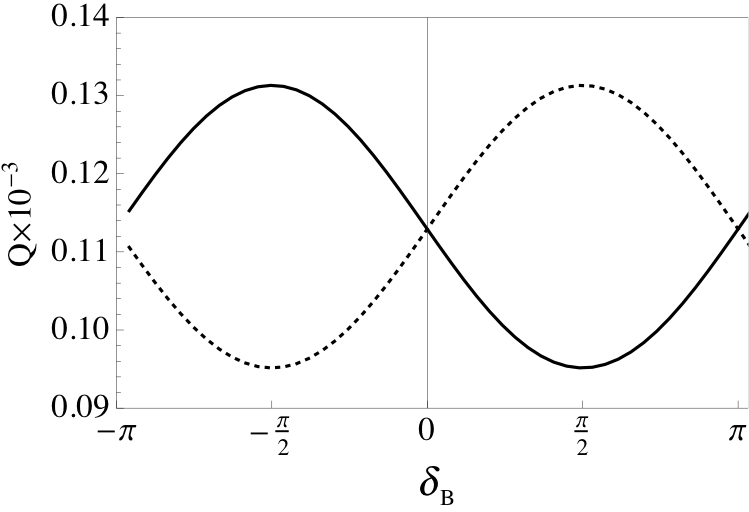}
\caption{$Q_{\rm{max}}(qrst)=Q(1010|1010)+Q(0101|1010)$ (dotted curve) and $Q_{\rm{max}}(qrst)=Q(1001|1010)+Q(0110|1010)$ (solid curve) are plotted vs $\delta_B$ for $\chi=0.24$, $\eta=0.04$, $\wp=1\times 10^{-5}$ and $\delta_A=\pi/2$. Figure reproduced from Fig.~4 of Khalique and Sanders~(2014)\cite{KS14}.
}
\label{fig:n3swaps}
\end{center}
\end{figure}

\begin{figure}[ht]
\begin{center}
\includegraphics[scale=0.6]{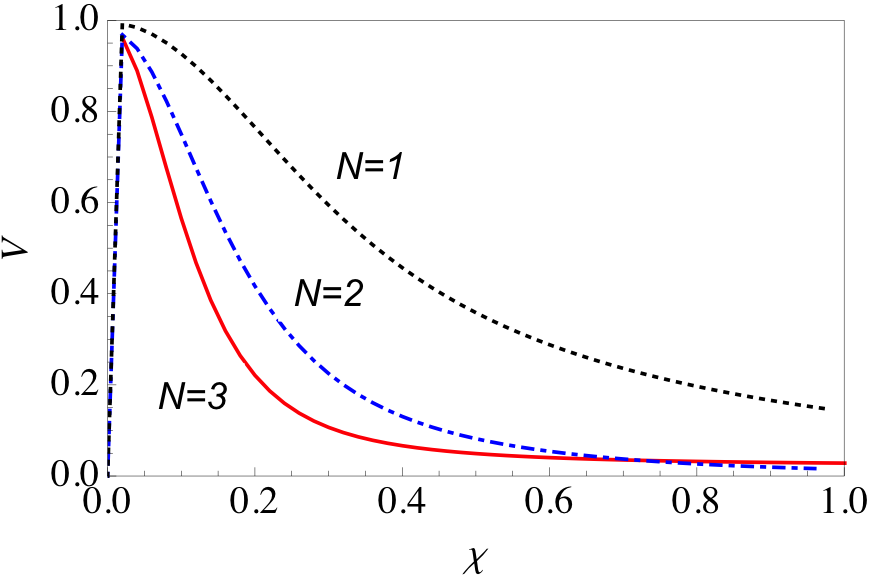}
\caption{Visibility is compared for $N=1$, $2$ and $3$ for varying $\chi$. Here, $\eta=0.04$, $\wp=1\times 10^{-5}$, $\delta_A=\delta_B=\pi/2$. Figure reproduced from Fig.~6 of Khalique and Sanders (2014)~\cite{KS14}.
}
\label{fig:comparisonchi}
\end{center}
\end{figure}

Here we present our results for visibility  and compare the same for $N=1$, $2$ and $3$ concatenated swaps~\cite{KS14}. In Fig.~\ref{fig:n3swaps} we give the coincidence probability $Q_{\rm{max}}(qrst)=Q(1010|1010)+Q(0101|1010)$ and $Q_{\rm{max}}(qrst)=Q(1001|1010)+Q(0110|1010)$ for varying $\delta_B$. The visibility calculated from the curve at $\delta_B=\pm\pi/2$ is $16\%$.
Visibility is compared for $N=1$, $2$ and $3$ swaps in Fig.~\ref{fig:comparisonchi}. 

\begin{figure}[ht]
\begin{center}
\includegraphics[scale=0.6]{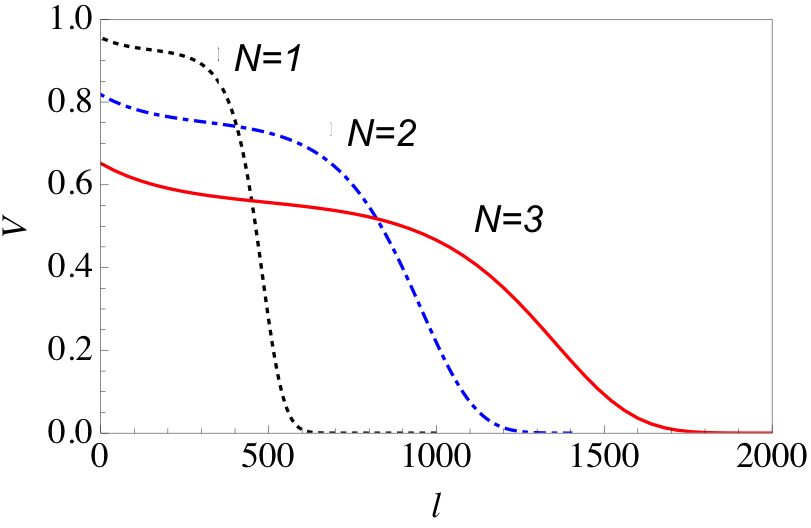}
\caption{Visibility is compared for $N=1$, $2$ and $3$ for various distances $l$ km. Here, $\chi=0.1$, $\eta_0=0.70$, $\wp=1\times 10^{-5}$, $\delta_A=\delta_B=\pi/2$, $\alpha=0.25$ dB/km and $\alpha_0=4$ dB. Figure reproduced from Fig.~6 of Khalique and Sanders (2014)~\cite{KS14}.
}
\label{fig:distcomparison}
\end{center}
\end{figure}

\begin{figure}[ht]
\begin{center}
\includegraphics[scale=0.6]{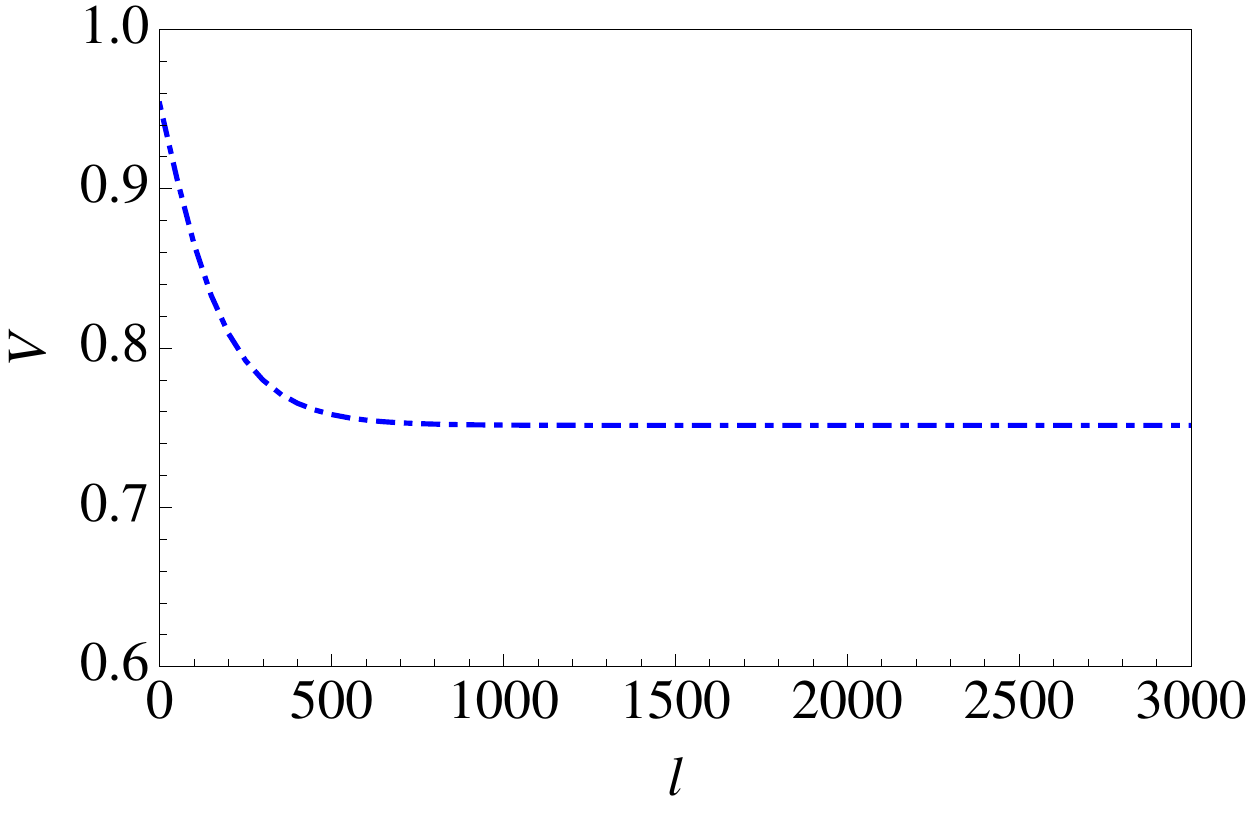}
\caption{Visibility vs distance, $l$ km, is shown for perfect detectors with $\eta_0=1$ and $\wp=0$ for $N=2$. Here, $\chi=0.1$, $\delta_A=\delta_B=\pi/2$, $\alpha=0.25$ dB/km and $\alpha_0=0$ dB.
}
\label{fig:perfectdetectors}
\end{center}
\end{figure}

The communication distance increases as the number of concatenations increases. Figure~\ref{fig:distcomparison} shows the comparison of visibility for $N=1$, $2$ and $3$ for various distances. The achievable distance increases to more than $1000$ km for $N=3$ but at the expense of very low visibility.  The increase in distance tends to saturate as the number of concatenations increases. The rapid fall off in visibility and limiting distance are due to detector dark counts and inefficiencies. For perfect detectors with $\eta_0=1$ and $\wp=1$, asymptotically large distance is achievable as shown in Fig.~\ref{fig:perfectdetectors}

\section{Long-distance Quantum Key Distribution protocol}
\label{sec:qkd}
\begin{figure}[ht]
\begin{center}
\includegraphics[scale=0.6]{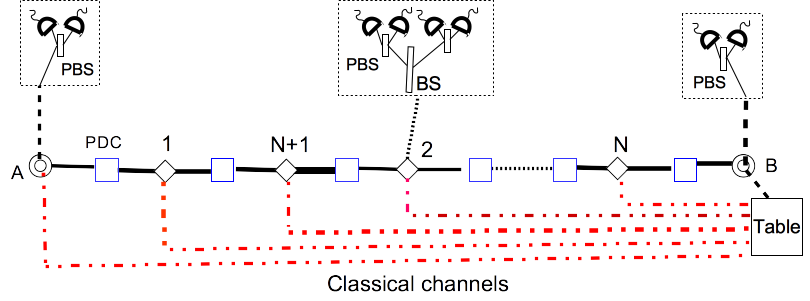}
\caption{Long-distance QKD setup is shown between users A and B. {\color{blue}$\square$} represents PDC source and $\diamond$ represents the BSM setup. Figure reproduced from Fig.~1 of Khalique and Sanders (2015)~\cite{KS15}.
}
\label{fig:longdistanceqkd}
\end{center}
\end{figure}

The concatenated entanglement swapping setup described above is implemented in long-distance QKD protocol~\cite{KS15}. The setup is shown in Fig.~\ref{fig:longdistanceqkd}. Two distant users A and B are connected by the concatenated entanglement swapping setup. Bell state measurements at the intermediate stations ensure entanglement at the two extreme ends. The results of two-photon coincidence at the intermediate stations are sent to B who calculates the visibility using these results and the two-photon coincidence at his and A's station by the formalism developed for concatenated swapping. The visibility is related to the quantum bit error rate (QBER)~\cite{GRT+02}
\begin{equation}
\rm{QBER}=\frac{1-V}{2}.
\end{equation}
The key-generation rate is 
\begin{equation}
  R  = R_\text{Shor-Preskill}R_\text{sifted}.
\end{equation}
Here, $R$ comprises the sifted key rate
\begin{equation}
\label{eq:sifted}
     R_\text{sifted}=\frac{1}{2}(\chi^2)^{2N}10^{(-\alpha l/{40N})4N}
     	(\eta^2/2)^{2N-1}\eta^2
\end{equation}
and the key retained after error correction and privacy amplification
\begin{equation}
\label{eq:shorpreskill}
        R_\text{Shor-Preskill}=1-\kappa H_2(Q)-H_2(Q).
\end{equation}
Here, $\kappa$ is the reconciliation efficiency, with $\kappa=1$ for perfect reconciliation. The net key rate is the product of the two rates,

We present the results obtained for maximized key generation rate $R_{\rm{max}}$ with optimum $\chi$, $\eta_0$ and $\wp$~\cite{KS15}. There is a trade off between $\eta_0$ and $\wp$, as for very high efficiency, the contribution of dark counts in detected photons also increases, which lowers the visibility.  We have used the trade-off corresponding to commonly used InGaAs detectors with
\begin{equation}
\wp=A\exp(B\eta_0)
\end{equation}
with typically
$A=6.1\times10^{-7}$ and $B=17$~\cite{CGR05}.

\begin{figure}[ht]
\begin{center}
\includegraphics[scale=0.425]{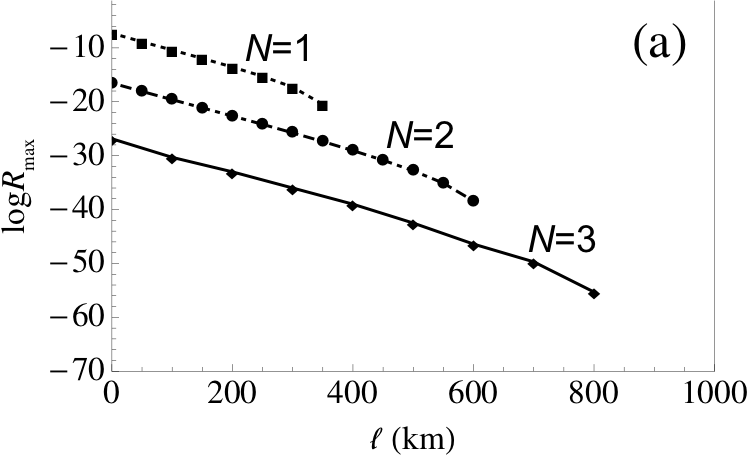}
	\includegraphics[scale=0.425]{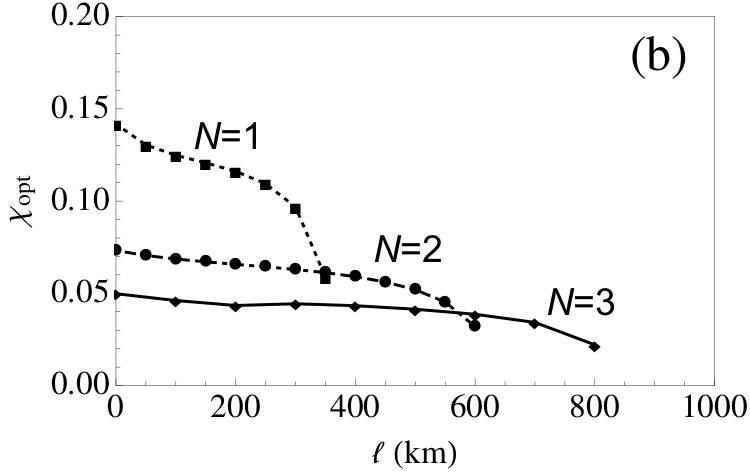}
	\includegraphics[scale=0.425]{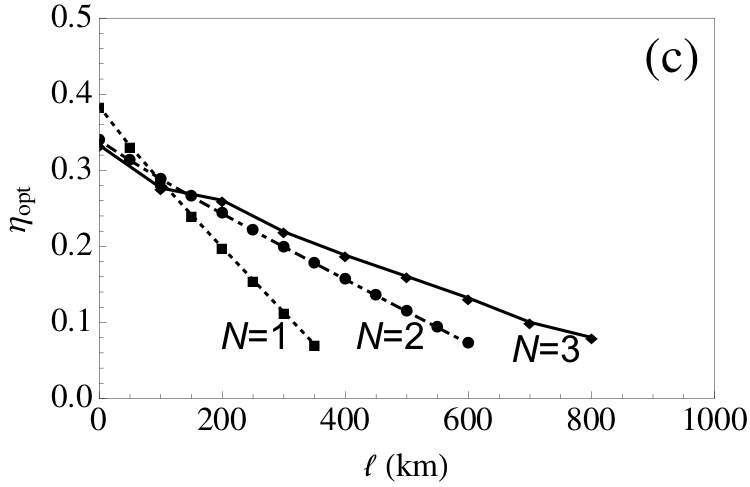}
\caption{(a) Plot of $\log R_{\rm{max}}$ vs distance $l$ km for $N=1$, $N=2$ and $N=3$. Corresponding optimal $\chi_{\rm{opt}}$ is shown in (b) and optimal efficiency $\eta_{\rm{opt}}$ is shown in (c). Here $\alpha=0.25$ dB/km and $\alpha_0=4$ dB. Figure reproduced from Fig.~3 of Khalique and Sanders (2015)~\cite{KS15}.
}
\label{fig:keyrate}
\end{center}
\end{figure}

\begin{figure}[ht]
\begin{center}
\includegraphics[scale=0.425]{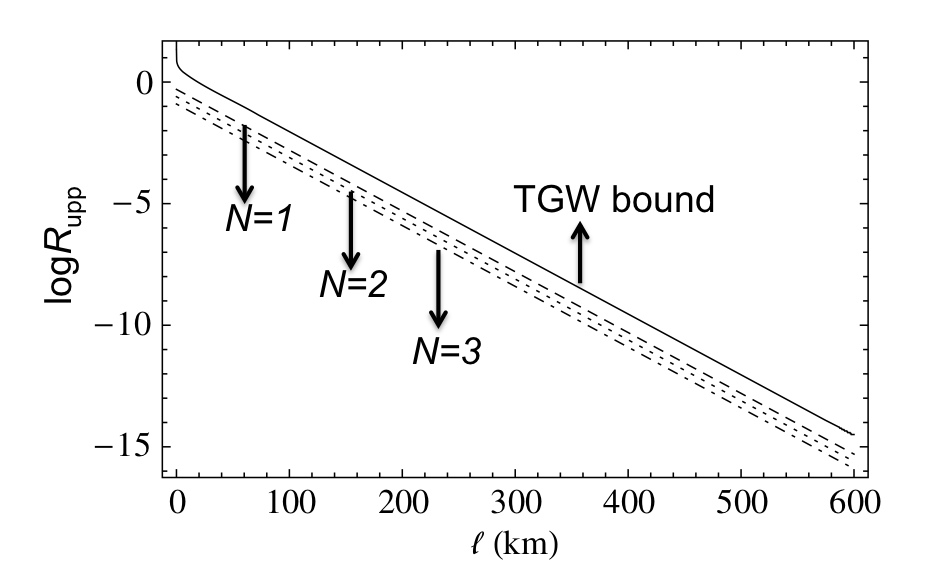}
\caption{Comparison between $R_{\rm{TGW}}$ bound and upper bound of Key rate for concatenated entanglement swapping for $N=1$, $N=2$ and $N=3$. Figure reproduced from Fig.~5 of  Khalique and Sanders (2015)~\cite{KS15}.
}
\label{fig:tgw}
\end{center}
\end{figure}

Maximum key generation rates $R_{\rm{max}}$ and the optimal $\chi$ and $\eta$ are shown in Fig.~\ref{fig:keyrate}. Distances upto~850 km are achievable for $N=3$ but at the cost of a very low key generation rate.  
We check the upper bound of key generation rate and compare it with Takeoka-Guha-Wilde (TGW) bound~\cite{TGW14}. The TGW bound gives an upper bound on key generation rate for non-repeater based QKD, which is
\begin{equation}
R_{\rm{TGW}}=\log_2\frac{1+10^{\frac{-\alpha l}{10}}}{1-10^{\frac{-\alpha l}{10}}}.
\end{equation}
The upper bound for the concatenated entanglement swapping set up is calculated by setting $R_{\rm{Shor-Preskill}}=1$ and thus $R=R_{\rm{sifted}}$.
The comparison in Fig.~\ref{fig:tgw} shows that the concatenated entanglement swapping key rates are very close to the TGW bound. Thus quantum memories are needed to further increase the key generation rates resulting from concatenated entanglement swapping setup.

\section{Conclusions}
\label{sec:conclusions}
We have presented our approach for calculation of visibility for concatenated entanglement swapping for arbitrary number of swaps and its application to long-distance QKD~\cite{KTS13, KS14, KS15}. Our model incorporates the practical resources. The results show that large distances can be achieved by concatenated entanglement swapping but this increase comes at the expense of attrociously low key generation rates. A trade off is needed between experiment run-time, resource parameters and key generation rates.

\section*{Acknowledgments} 
We thank valuable discussions with Wolfgang Tittel, Michael Lamoreux, Artur Scherer, Norbert L{\"u}tkenhaus and Pengqing Zhang and financial support from AITF, NSERC and 1000 Talent Plan of China. This research has been enabled by the use of computing resources provided by WestGrid and Compute/Calcul Canada.

\bibliography{report} 

\begin{thebibliography}{10}

\bibitem{GRT+02}
Gisin, N., Ribordy, G., Tittel, W., and Zbinden, H., ``Quantum cryptography,''
  {\em Rev. Mod. Phys.}~{\bf 74},  145--195 (Mar 2002).

\bibitem{TYC+14}
Tang, Y.-L., Yin, H.-L., Chen, S.-J., Liu, Y., Zhang, W.-J., Jiang, X., Zhang,
  L., Wang, J., You, L.-X., Guan, J.-Y., Yang, D.-X., Wang, Z., Liang, H.,
  Zhang, Z., Zhou, N., Ma, X., Chen, T.-Y., Zhang, Q., and Pan, J.-W.,
  ``Measurement-device-independent quantum key distribution over 200 km,'' {\em
  Phys. Rev. Lett.}~{\bf 113},  190501 (Nov 2014).

\bibitem{GEN+16}
Gleim, A.~V., Egorov, V.~I., Nazarov, Y.~V., Smirnov, S.~V., Chistyakov, V.~V.,
  Bannik, O.~I., Anisimov, A.~A., Kynev, S.~M., Ivanova, A.~E., Collins, R.~J.,
  Kozlov, S.~A., and Buller, G.~S., ``Secure polarization-independent
  subcarrier quantum key distribution in optical fiber channel using bb84
  protocol with a strong reference,'' {\em Opt. Express}~{\bf 24},  2619--2633
  (Feb 2016).

\bibitem{GT07}
Gisin, N. and Thew, R., ``Quantum communication,'' {\em Nature Photon.}~{\bf
  1},  165--171 (03 2007).

\bibitem{KGD+16}
Krovi, H., Guha, S., Dutton, Z., Slater, J.~A., Simon, C., and Tittel, W.,
  ``Practical quantum repeaters with parametric down-conversion sources,'' {\em
  Appl. Phys. B}~{\bf 122}(3),  1--8 (2016).

\bibitem{KTS13}
Khalique, A., Tittel, W., and Sanders, B.~C., ``Practical long-distance quantum
  communication using concatenated entanglement swapping,'' {\em Phys. Rev.
  A}~{\bf 88},  022336 (Aug 2013).

\bibitem{KS14}
Khalique, A. and Sanders, B.~C., ``Long-distance quantum communication through
  any number of entanglement-swapping operations,'' {\em Phys. Rev. A}~{\bf
  90},  032304 (Sep 2014).

\bibitem{KS15}
Khalique, A. and Sanders, B.~C., ``Practical long-distance quantum key
  distribution through concatenated entanglement swapping with parametric
  down-conversion sources,'' {\em J. Opt. Soc. Am. B}~{\bf 32},  2382--2390
  (Nov 2015).

\bibitem{BBM92}
Bennett, C.~H., Brassard, G., and Mermin, N.~D., ``Quantum cryptography without
  {B}ell's theorem,'' {\em Phys. Rev. Lett.}~{\bf 68},  557--559 (Feb 1992).

\bibitem{RR06}
Rohde, P.~P. and Ralph, T.~C., ``Modelling photo-detectors in quantum optics,''
  {\em J. Mod. Opt.}~{\bf 53}(11),  1589--1603 (2006).

\bibitem{SHST09}
Scherer, A., Howard, R.~B., Sanders, B.~C., and Tittel, W., ``Quantum states
  prepared by realistic entanglement swapping,'' {\em Phys. Rev. A}~{\bf 80},
  062310--062329 (Dec 2009).

\bibitem{CGR05}
Collins, D., Gisin, N., and de~Riedmatten, H., ``Quantum relays for long
  distance quantum cryptography,'' {\em J. Mod. Optic.}~{\bf 52}(5),  735--753
  (2005).

\bibitem{TGW14}
Takeoka, M., Guha, S., and Wilde, M.~M., ``Fundamental rate-loss tradeoff for
  optical quantum key distribution,'' {\em Nat. Commun.}~{\bf 5} (10 2014).

\end{thebibliography}

\end{document}